# Imaging inspired characterization of single photons carrying orbital angular momentum

V<small>IMLESH</small> K<small>UMAR</small>,[1,*] V<small>ARUN</small> S<small>HARMA</small>,[1,2] S<small>ANDEEP</small> S<small>INGH</small>,[1,3] S. C<small>HAITANYA</small> K<small>UMAR</small>,[4] A<small>NDREW</small> F<small>ORBES</small>,[5] M. E<small>BRAHIM</small>-Z<small>ADEH</small>,[4,6] G. K. S<small>AMANTA</small>[1]

[1]*Photonic Sciences Lab, Physical Research Laboratory Ahmedabad, Navrangpura, Gujarat 380009, India*
[2]*Institute of Physics, University of Graz, NAWI Graz, Universitätsplatz 5, 8010 Graz, Austria*
[3]*Indian Institute of Technology-Gandhinagar, Ahmedabad, 382424, Gujarat, India*
[4]*ICFO-Institut de Ciencies Fotoniques, The Barcelona Institute of Science and Technology, 08860 Castelldefels (Barcelona), Spain*
[5]*School of Physics, University of the Witwatersrand, Private Bag 3, Johannesburg 2050, South Africa*
[6]*Institucio Catalana de Recerca i Estudis Avancats (ICREA), Passeig Lluis Companys 23, Barcelona 08010, Spain*

*\*Corresponding author: vimlesh@prl.res.in*

**We report on an imaging inspired measurement of orbital angular momentum (OAM) using only a simple tilted lens and an Intensified Charged Coupled Device (ICCD) camera, allowing us to monitor the propagation of OAM structured photons over distance, crucial for free-space quantum communication networks. We demonstrate measurement of OAM orders as high as $l_s$=14 in a heralded single-photon source (HSPS) and show, for the first time, the imaged self-interference of photons carrying OAM in a modified Mach-Zehnder Interferometer (MZI). The described methods reveal both the charge and order of a photon's OAM, and provide a proof of concept for the interference of a single OAM photon with itself. Using these tools, we are able to study the propagation characteristics of OAM photons over distance, important for estimating transport in free-space quantum links. By translating these classical tools into the quantum domain, we offer a robust and direct approach for the complete characterization of a twisted single photon source, an important building block of a quantum network.**

Quantum entanglement has emerged as a platform for new applications in the field of communication [1], computation [2], and imaging [3]. For such applications, there has been significant effort to develop reliable single-photon sources based on various methods such as trapped atoms/ions [4], nitrogen vacant diamond [5], and spontaneous parametric down-conversion (SPDC) in second-order nonlinear media [6]. In this regard, the efficient generation of heralded single-photon sources (HSPS) through the SPDC process under ambient conditions has emerged as a platform for applications in quantum communication [7], teleportation, and quantum imaging [8,9]. Single-photon sources utilized in such applications exploit polarization as a two-dimensional basis for entangled pairs of photons. But to utilize such sources to their full potential, there is a need to extend the two-dimensional basis by entangling photon pairs in another degree of freedom such as space [10] or frequency [11]. Of the various possible options, spatial properties such as orbital angular momentum (OAM) [12] have emerged as viable solutions with relative ease of implementation. Since OAM is conserved in the SPDC process, this leads to a high correlation in the biphoton state generated in HSPS. Photons with OAM are characterized by helical phase front, $exp(il\phi)$, where $(l\phi)$ is the phase variation in the transverse plane with phase singularity at the center of the beam, and $l$ is known as the topological charge (order) of the beam. This azimuthal phase dependence can be found in many spatial modes, for example, the Laguerre-Gaussian (LG) and Bessel-Gaussian (BG) modes, forming an infinite-dimensional Hilbert space based on structured single and entangled photons [13,14]. The availability of an infinite number of orthogonal states makes photons with OAM a suitable candidate for classical and quantum communication. Since photons with OAM are also interpreted as a twist in the wavefront along the propagation direction, they are so-called twisted photons [14]. With the ever-increasing applications of twisted single-photon sources, it is essential to devise new techniques for complete characterization of such states.

In the classical domain, various techniques have been developed to determine the spatial properties of beams carrying OAM [15]. On the other hand, the measurement of OAM content and correlation in bi-photon states is performed using a phase flattening technique with the help of spatial light modulators (SLMs) and a single-mode fiber. In this technique, the contribution of different OAM modes of the beam is measured by projecting the beam in fundamental Gaussian mode into a single-mode fiber [13]. This projection scheme results in the loss of the spatial characteristics of such propagating OAM modes. Recent advances in the field of detection of correlated photon pairs using intensified charge-coupled device (ICCD) is attracting the attention of the scientific community. ICCD-based detection has proved to be versatile for real-time imaging of entangled states [16], measuring higher-order entangled states [17], and imaging with a low number of photons [18]. Direct detection of free-space propagating single photons makes the ICCD a preferred tool to fully characterize spatial modes. Further, direct imaging of propagating modes on ICCD can help utilize simple techniques established in the classical domain for OAM detection and propagation characteristics measurements. In classical optics, the tilted lens technique [19] has evolved as one of the simplest methods to implement over a broad wavelength range. In this technique, the beam with OAM mode, $l$, after passing through a convex lens tilted in the transverse plane with the optic axis, splits into $|l|$ +1 number of lobes at the back focal plane of the lens. Due to the simplicity of the technique, it has found great interest for the detection of OAM content in the mid-infrared [20] and ultraviolet [21]. Therefore, the novel combination of the tilted lens technique and coincidence imaging in ICCD can potentially open up the possibility of measuring the evolution of single photon OAM modes during propagation, without perturbing the experimental setup. At present, this is not possible with conventional projective measurements, because of the imaging constraints from fiber to crystal. To circumvent this, a recent demonstration used diffraction off a triangular aperture for quantifying the OAM content in HSPS, but this approach was limited to low values of OAM ($l$= 2) [22].

Here, we overcome these restrictions and demonstrate a robust, easy-to-implement approach based on the tilted lens technique for measuring the OAM content of single photons. Based on this technique, we demonstrate the measurement of OAM content of a single photon with a maximum value of up to 14. To verify the reliability of the technique, we measure the single-photon self-interference [23] for the twisted photons with OAM mode as high as 7. We also fully characterize the propagation dynamics of the single twisted photons and observe the single-photon divergence with its OAM modes similar to the classical vortex beams. Using the ICCD in coincidence imaging architecture, the spatial properties of the twisted photons are measured, for the first time to the best of our knowledge. This work provides an excellent platform for the design of future sources and detectors, such as telescopes to collect higher-order OAM modes for free-space quantum communication with high dimensional entangled states.

The schematic of the experimental setup for the generation of HSPS in higher-order spatial modes is shown in Fig. 1. A single-mode continuous-wave laser (Toptica, *topmode*) of spectral width of 12 MHz centered at 405 nm and providing an output power of 100 mW is used as the pump source. A combination of a half-wave plate ($\lambda/2$) and a polarizing beam-splitter (PBS1) is used for systematic control of pump power to the experiment. To generate the bi-photon OAM state, we use the method of pump beam OAM transfer to the signal and idler photons [24]. This technique has the advantage of higher efficiency over the conventional method of imprinting OAM onto single photons after generation. The SLM (Hamamasu *LCOS*) transforms the Gaussian pump beam into different optical vortices. The first-order diffracted beam of the SLM containing the OAM mode is extracted from the remaining beams after the 50:50 beam-splitter (BS1) and focused at the center of the nonlinear crystal using a lens (L1) of focal length, $f$=100 mm. A 30-mm-long periodically-polled KTP (PPKTP) crystal with 2x1 mm$^2$ aperture incorporating a single grating period of $\Lambda$=10 μm is used as the nonlinear gain medium for type-II ($e{\to}oe$) quasi-phase-matched parametric down-conversion of 405 nm into degenerate photon pairs at 810 nm. The $\lambda/2$-plate before the lens, L1, controls the polarization of the pump beam for optimum phase-matching. The crystal is housed in an oven for temperature control to ensure the generation of collinear, degenerate paired photons (signal and idler) at 810 nm in orthogonal polarization states, say signal (H-polarized) and idler (V-polarized). The generated photon pairs are collimated using a lens (L2) of focal length, $f$=200 mm, and extracted from the pump using the notch filer (NF) at 405 nm. A polarizing beam-splitter (PBS2) separates the signal and idler in the transmission and reflection ports, respectively. In the current scheme, we performed three different experiments marked with different color codes in Fig. 1. We guided the photons in various combinations using different flip mirrors (FM1, FM2, FM3), in order to perform Hong-Ou-Mandel (HOM) interferometry [25,26], coincidence imaging, and Mach-Zehnder interferometry (MZI).

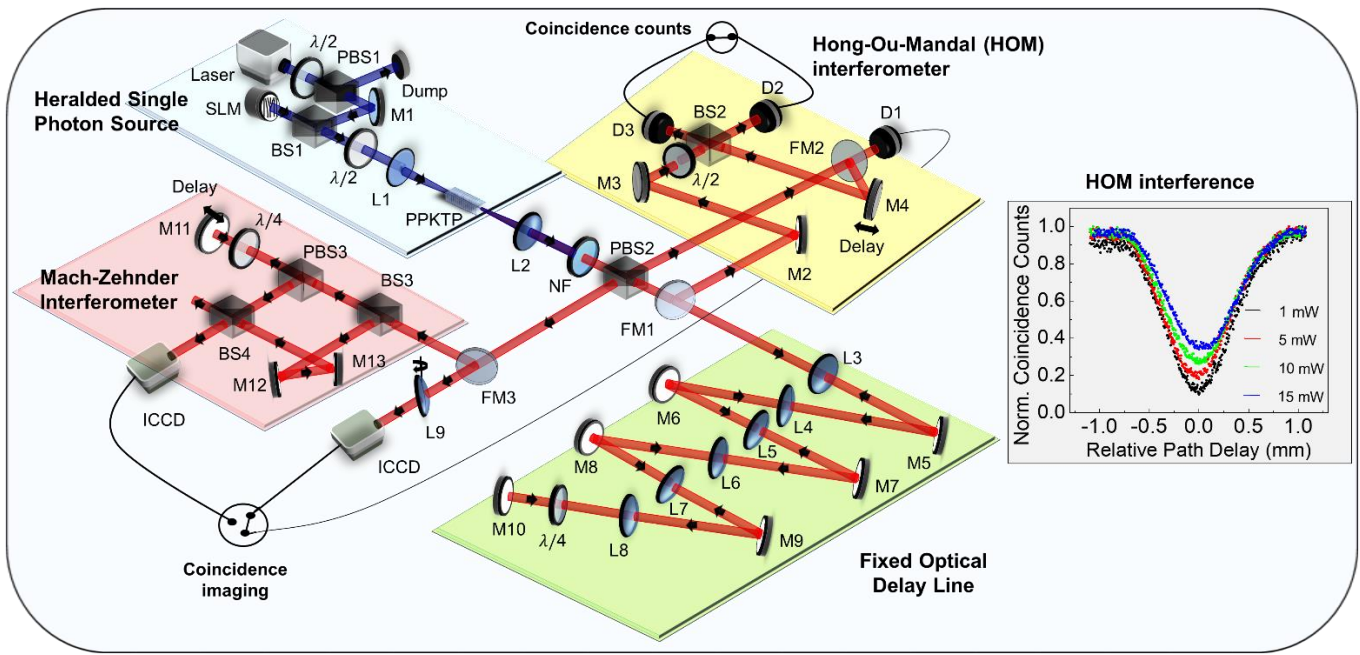

Fig.1. Schematic of the experimental setup for HSPS. M1-12: Mirrors, PBS1-3: Polarizing beam splitters, BS1-4: Beam splitters, SLM: Spatial light modulator, L1-9: Lens, NF: Notch filter, FM: Flip mirror, D1-3: Single-mode fiber coupled single photon counting module (SPCM) with interference filter, IF of bandwidth ~3 nm centered at 810 nm (not shown), ICCD: Intensified charged coupled device. Inset: HOM interference - variation of coincidence counts between the two output ports of the beam-splitter of the HOM interferometer as a function of relative path delay between the paired photons generated through the type-II SPDC process for pump powers of 1, 5, 10, and 15 mW.

First, we studied the single-photon nature of the down-converted output by guiding the paired photons, using two flip mirrors, FM1 for signal and FM2 for idler, into the HOM interferometer comprising three mirrors (M2, M3, M4) on a translation stage and a beam-splitter (BS2). The motorized translation stage has a resolution of ~0.05 µm and a travel range of 25 mm. Since the paired photons are in orthogonal states, the λ/2-plate placed before BS2 is used to convert both photons in the same polarization state. Using the single-photon counting modules (SPCM), D2 and D3 connected with single-mode fiber, and an interference filter (IF) of ~3 nm bandwidth centered at 810 nm and time-to-digital converter, we measured the coincidence counts between the two output ports of BS2. Keeping the crystal temperature at 39 °C for collinear SPDC process, we varied the relative time delay between the paired photons through the path delay by translating the mirror, M4, and recorded the coincidence counts. The results are shown in the inset of Fig. 1. The maximum coincidence counts increase with pump power, but at the expense of visibility. As evident from the plot, the normalized coincidence counts measured with a coincidence window of 5 ns, for pump powers of 1, 5, 10, and 15 mW, vary with the path delay between the photons, resulting in HOM dip with uncorrected HOM visibilities of 0.90, 0.82, 0.737, and 0.66, respectively. Using theoretical fit to the experimental results, we measured the full-width at half-maximum (FWHM) of the HOM dip to be about 760 µm, corresponding to the spectral width of the single photons of ~0.85 nm, much smaller than the bandwidth (~3 nm) of the IF used in the experiment. In fact, by removing the IF from the detectors, we have observed similar HOM characteristics and HOM dip width. Therefore, we removed the IF from the ICCD camera in the coincidence imaging experiments to avoid photon loss due to the transmission loss of the filter. On the other hand, single-photon imaging requires the integration of a large number of frames, and the number of frames reduces with the increase in the brightness of the source. Therefore, we used a pump power of 10 mW, while ensuring the single-photon nature of the source with high coincidence counts and high HOM visibility for all experiments in the current report.

Confirming the single-photon characteristics of our HSPS, we studied the OAM transfer of the pump beam to the down-converted photons through coincidence imaging. Removing the flip mirrors (FM1, FM2, FM3), we detected the V-polarized idler photon reflected off PBS2 by the SPCM (D1), and triggered the ICCD camera in coincidence imaging mode to measure the spatial distribution of the partner H-polarized signal photon transmitted through PBS2. However, ICCD has an unwanted electronic delay of ~80 ns. Therefore, to detect the correct partner photon, we delayed the signal photon by a physical length of around 24 m using the fixed delay line. The delay line comprises six biconvex lenses (L3-L8) of focal length, $f$=1000 mm, placed in $4f$ imaging configuration corresponding to a path delay of 12 m in single pass. Six plane mirrors (M5-M10) are used to fold the beam to accommodate the delay line on the optical table. The signal is then retro-reflected using M10 through a λ/4-plate, resulting in a round-trip path delay of 24 m, compensating the 80 ns time delay for triggering the ICCD. The presence of λ/4-plate flips the polarization state of the signal from horizontal to vertical and is subsequently reflected by PBS2 to ICCD camera for coincidence imaging, as shown in Fig. 1. By virtue of OAM conservation, the pump OAM mode, $l_p$, can be distributed between signal and idler photons in different possible combinations of OAM modes ($l_s, l_i$). However, the detection of idler photon through the single-mode fiber restricts the idler photon to carry OAM mode, $l_i$=0, of Gaussian mode. As a result, the pump OAM mode, $l_p$, is automatically transferred to the correlated signal photon, $l_s=l_p$, which is subsequently imaged by the ICCD operated in coincidence imaging mode.

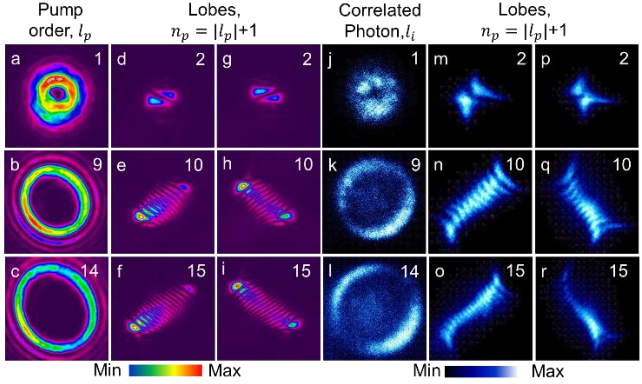

Fig. 2. (a-c). Pump OAM intensity spatial distribution of order, $|l_p|$=1, 9, 14. Lobe structure intensity distribution having ($|l_p|$+1) number of bright lobes for pump OAM, (d-f) $l_p$ =1, 9, 14 and (g-i) $l_p$ = -1, -9, -14 recorded using CCD in through the tilted lens technique. Spatial distribution of (j-l) signal photons of OAM modes, $|l_s|$=1, 9, 14 and corresponding lobe structure having ($|l_s|$+1) number of bright lobes for signal photon OAM mode, (m-o) $l_s$ =+1, +9, +14, and (p-r) $l_s$ = -1, -9, -14, recorded using coincidence imaging using ICCD.

To confirm the pump beam carrying OAM mode, we imaged the pump beam after mode conversion using the SLM, with the results shown in Fig. 2. As evident from the first column, (a-c), the pump beam has doughnut-shape intensity distribution with increasing dark core area resembling vortex beam of different OAM modes. To confirm the presence of OAM mode in the pump beam, we used the simple titled lens technique [19], where the pump beam is passed through a biconvex lens tilted in the transverse plane and subsequently recorded at the back focal plane. The tilt of the lens introduces astigmatism in the input beam and splits the vortex beams into characteristic lobe structure with lobe number, $n=|l|+1$, for vortex beam of order, $\pm l$. While the number of lobes provides information about the order of OAM, the orientation of the bright lobes contains information about the sign of the OAM mode. As evident from the second column, (d-f), of Fig. 2, the pump beam splits into 2, 10, and 15 lobes oriented along +45°, confirming the pump OAM mode of order, $|l_p|$ = 1, 9, and 14, respectively. Similarly, we also measured the OAM content of the pump beam by flipping the hologram on the SLM. While the intensity distribution of the beams remains unchanged, the lobe structures shown in the third column, (g-i), of Fig. 2 are oriented at -45° with the same number of lobes, confirming the order of the pump OAM mode, $|l_p|$ = 1, 9, and 14, but in the opposite sign. Knowing the OAM content of the pump beam and phase winding direction (the sign of the OAM mode), we measured the coincidence images of the signal triggered with the idler photon, with the results shown in the fourth column, (j-l), of Fig. 2. As evident, the coincidence images recorded by accumulating 30, 120, and 430 frames, each of having 1 s of exposure time for pump OAM modes, $l_p$ = ±1, ±9, and ±14, respectively, at fixed power, have clear doughnut-shaped spatial distribution with increasing dark core size similar to the pump beams. To verify the OAM mode of the signal photon, we used a lens (L9) of focal length, $f$=1000 mm, tilted in the transverse plane, and recorded the coincidence images by accumulating 30, 120, and 650 number of frames. Here we kept the ICCD camera at the focal plane of the lens and ensured a total path delay of more than 24 m. As evident from the fifth column, (m-o), and sixth column, (p-r), of Fig. 2, the doughnut-shaped spatial distribution of the signal split into characteristic lobes oriented at +45° and -45°, respectively, similar to the pump beam. Count of the number of lobes shows that the signal photon carries OAM of order, $|l_s|$=1, 9, and 14, same as the pump OAM mode, confirming the OAM conservation in type-II SPDC process ($|l_p|=|l_s|+|l_i|$, $|l_i|$=0). Orientation of the lobe structure also confirms that signal OAM mode is of the same sign as the pump OAM mode. These measurements confirm the successful transfer of the OAM mode of the pump to the single photons and the implementation of a simple approach to characterizing the OAM content even at the single-photon level.

Further, to confirm the presence of the OAM at the single-photon level, we performed self-interference measurement for the single photons in a modified Mach-Zander Interferometer (MZI), as shown in Fig. 1. Using a flip mirror (FM3), we coupled the single photons into the modified MZI, comprising two 50:50 beam-splitters (BS3, BS4), a polarizing beam-splitter (PBS3) cube, three plane mirrors (M11-M13), and a λ/4-plate. The signal photon carrying OAM of order, $l_s$, splits into two paths with equal probability by BS3 and subsequently recombined by BS4 after propagating through different paths. However, the combination of mirror, M11, placed on a translation stage, λ/4-plate, and PBS3, ensures equal path length between the arms of the MZI. The coincidence images of the signal photon detected by the ICCD camera placed at one of the output ports of BS4 are shown in Fig. 3. Using the pump beam OAM mode of order, $l_p$=±1, ±4, and ±7, we blocked one of the arms of the MZI and recorded the spatial profile in a doughnut shape with increasing dark core size, with the results shown in the first column, (a-c), of Fig. 3. As evident, the heralded signal photon carries doughnut-shaped spatial distribution with increasing dark core size with the increase of the OAM mode of pump beam from $l_p$=±1 to ±7. To confirm the presence of phase singularity, we added a nominal lateral displacement and tilt to one of the arms of the modified MZI with respect to another arm and recorded the off-axis interference with fork-like intensity distribution [23] on the ICCD camera. From interference patterns shown in the second column, (d-f), and third column, (g-i), of Fig. 3, for the pump OAM mode, $l_p$=1, 4, and 7, and $l_p$=-1, -4, and -7, respectively, we observe the fork pattern of the vortex-vortex interference of same order and sign, as expected from the geometry of the MZI (even number of reflections in one arm and odd reflections in another arm). While the fork pattern is observed (see the red circles) for lower OAM orders, it is difficult to see at higher OAM orders due to the larger dark core size. However, we observe the difference between the number of fringes in the upper and lower region of the images (marked by the red circles) to be equal to $2l_p$ for a given pump OAM mode, $l_p$. From the OAM conservation, $|l_p| = |l_s| + |l_i|$, and $l_i$=0 (Gaussian mode due to the single-mode fiber), we can confirm the signal photons to carry the pump OAM mode, $|l_s| = |l_p|$. The change in the orientation of the fork patterns confirms the sign of the OAM mode of the single photon with the change in the sign of the pump OAM mode.

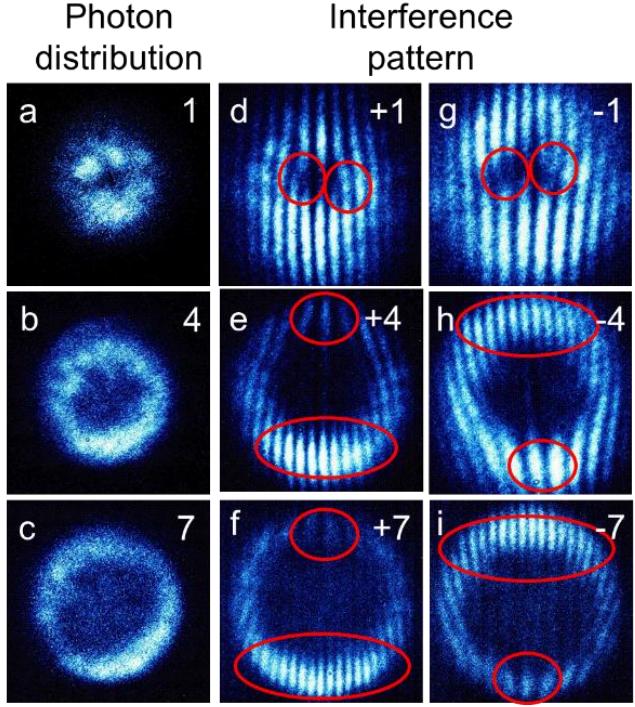

Fig. 3. (a-c) Spatial intensity distribution of single photons generated from the pump OAM mode, $|l_p|$=1, 4, and 7. Self-referenced interference of signal photon OAM of order, (d-f) $l_s$=1,4,7, and (g-i) $l_s$=-1,-4,-7. All the images recorded through the coincidence imaging in the ICCD.

Knowing the OAM content of the single photon directly transferred from the classical pump beam through the nonlinear interaction, we then proceeded to study the propagation characteristics of the single-photon carrying OAM mode. In doing so, we removed the flip mirror, FM3, changed the focal length of the lens, L9, to $f$= 500 mm, and detected the single photon using the ICCD camera placed at a distance of 10 cm from the focal plane of the lens. Changing the pump OAM mode, $|l_p|$=1, 3, 5, at a constant power of 10 mW, we recorded the ring-shaped intensity distribution of the signal photon at an interval of 5 cm over a total distance of 35 cm in the propagation direction, $z$, considering the focal plane of L9 as $z$=0. The results are shown in Fig. 4. Using the recorded images, we plotted the line intensity profile along the diameter of the doughnut image of the signal photons and measured the diameter of the doughnut ring with the maximum intensity along with propagation distance for pump OAM modes, $|l_p|$ = 1, 3, and 5. The results are shown in Fig. 4(a). As evident, the diameter of the single-photon doughnut spatial profile corresponding to OAM modes, $|l_s|$=$|l_p|$=1, 3, and 5, increases with propagation distance, $z$, and the rate of increase is higher for higher-order OAM modes. These results confirm divergence of single photons with OAM order and propagation. However, to gain a better understanding about the single-photon divergence, we measured the beam divergence of the pump beam of OAM modes, $|l_p|$ = 1, 3, and 5, under a similar experimental condition, with the results shown in Fig. 4(b). As reported previously [27], here we also observe, as shown in Fig. 4(b), the increase in the diameter of the pump vortex beam measured in reference to the maximum intensity points (same as the inner and outer radius) with the propagation distance and OAM order. Using the mathematical expression for vortex beam radius with propagation distance [27] and fitting to the experimental results of Fig. 4(a) and 4(b), we calculate the value of $w_0 \times r(0)$ and $z_R^2$, where $w_0$ is the beam waist at $z = 0$, $z_R$ is Rayleigh length, and $r(0)$ is the ring radius at $z$=0. Given the unavailability of suitable formula for the waist size of vortex beams, we used the same terminology [27] to calculate the vortex beam divergence, that is the rate of change of beam radii with propagation distance. Using the values of $w_0 \times r(0)$ and $z_R^2$ from Fig. 4(a) and 4(b) for different orders, we experimentally (black dots) determined the divergence of single-photon and classical pump beams of different OAM modes, with the results shown in Fig. 4(c) and 4(d), respectively. Using the relevant expressions [27], we also calculated the theoretical variation of divergence (red dots) with OAM order for single-photon and classical pump beams, confirming close agreement with experimental results. As evident from Fig. 4(c) and 4(d), the divergence of single-photon and the pump beam increases linearly with its OAM content. We find the ratio of divergence value of single-photon OAM mode to the corresponding pump beam OAM mode to be almost constant. Therefore, for free-space propagation of single photons with OAM modes, one can measure the divergence properties of the classical pump beam to estimate the divergence of the single photons without directly measuring the same for the single photons using cumbersome experimental architecture.

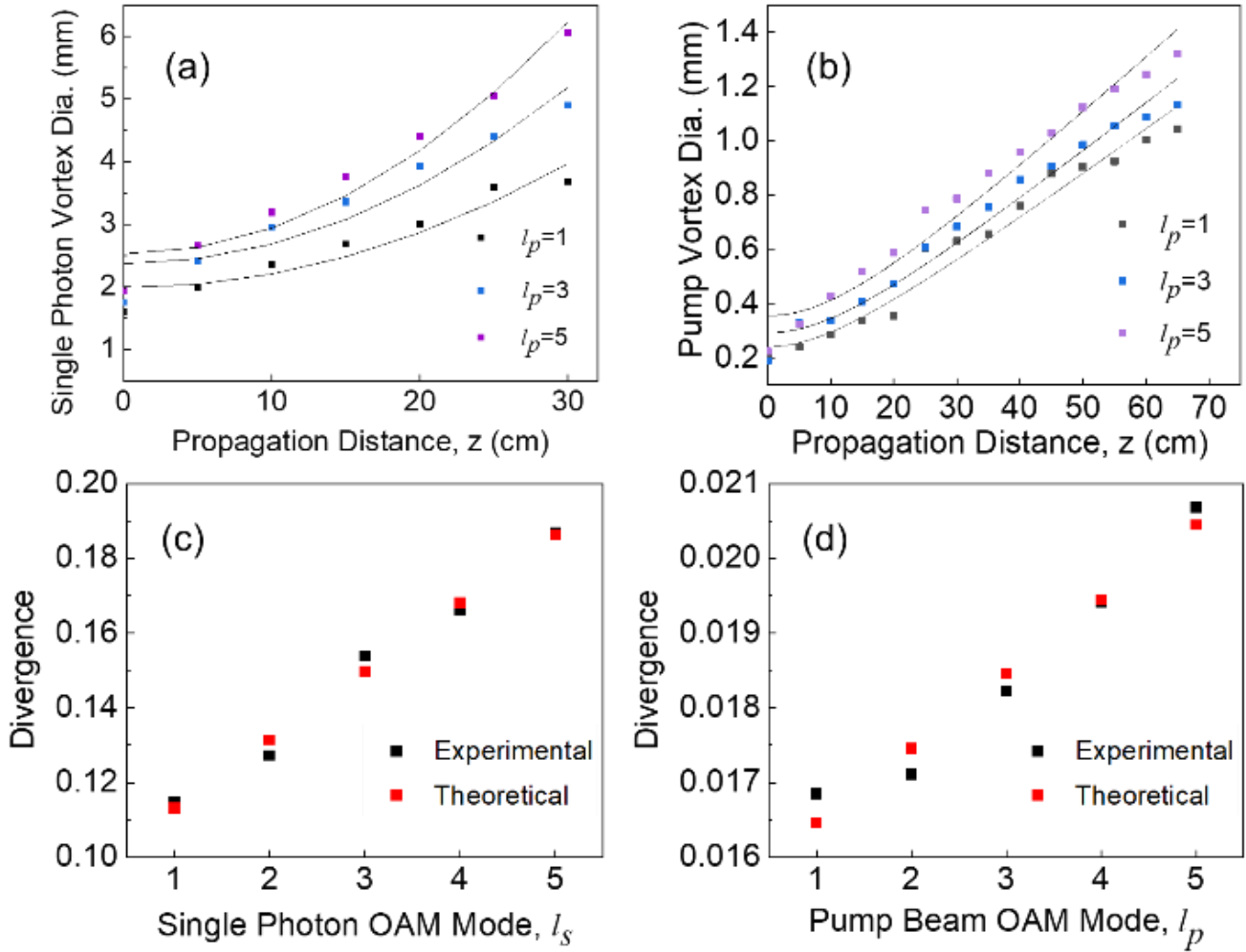

Fig. 4. (a) Variation of the diameter of doughnut spatial distribution of single-photon and (b) corresponding pump vortex beam of OAM mode, $l_p$ = 1, 3, and 5 with propagation distance. Divergence variation of (c) single-photon and (d) corresponding pump beam as a function of OAM mode. Solid lines are a guide to the eyes.

In conclusion, we have successfully used an imaging inspired setup to measure the OAM of a signal photon using a simple tilted convex lens. In addition to the effectiveness of this simple technique for OAM determination, we have used modified MZI and visualized the interference nature of photon-carrying OAM for $l_s$ = 1 to 7 for both positive and negative OAM charges. We have also determined the divergence values of the OAM modes for both single photon (810 nm) and classical pump beam (405 nm) for $l_p$= 1 to 5 and found the divergence nature for both pump OAM and photon OAM to remain almost similar. Therefore, one can simply measure the divergence characteristics of the pump beam and estimate the divergence of the single photons, as required for quantum communication with dimensional states. The complete characterization will provide a strong platform for future studies involving higher-order spatial modes in HSPS.

**Disclosures**. "The authors declare no conflicts of interest."

## Full References